# Direct observation of bubble-assisted electroluminescence in liquid xenon

E. Erdal[1*], L. Arazi[1], V. Chepel[2], M. L. Rappaport[1], D. Vartsky[1], and A. Breskin[1]

[1]*Department of Particle Physics and Astrophysics*
*Weizmann Institute of Science, Rehovot 7610001, Israel*
[2]*Department of Physics, University of Coimbra, 3004-516 Coimbra, Portugal*
E-mail: eran.erdal@weizmann.ac.il

ABSTRACT:

Bubble formation in liquid xenon underneath a Thick Gaseous Electron Multiplier (THGEM) electrode immersed in liquid xenon was observed with a CCD camera. With voltage across the THGEM, the appearance of bubbles was correlated with that of electroluminescence signals induced by ionization electrons from alpha-particle tracks. This confirms recent indirect evidence that the observed photons are due to electroluminescence within a xenon vapor layer trapped under the electrode. The bubbles seem to emerge spontaneously due to heat flow from 300K into the liquid, or in a controlled manner, by locally boiling the liquid with resistive wires. Controlled bubble formation resulted in energy resolution of $\sigma/E \approx 7.5\%$ for ~6,000 ionization electrons. The phenomenon could pave ways towards the conception of large-volume 'local dual-phase' noble-liquid TPCs.

---

[*] Corresponding author

# Contents

## 1  Introduction

The concept of Liquid Hole-Multipliers (LHMs) was introduced in [1] as a potential method for detection of ionization electrons and scintillation photons in future large-volume noble-liquid time projection chambers (TPCs). The idea was to employ LHMs in a *single-phase* noble liquid TPC scheme, which should be simpler to scale to large dimensions than the dual-phase design used in current dark matter experiments [2] [3] [4]. The original LHM concept comprised a cascade of perforated electrodes (e.g. Thick Gas Electron Multipliers (THGEMs) [5] or Gas Electron Multipliers (GEMs) [6]) immersed in the noble liquid, with CsI photocathodes deposited on their surfaces. It was suggested that electrons focused into the electrodes' holes (either ionization electrons drifting towards the LHM or photoelectrons induced by primary scintillation on the photocathode), would lead to electroluminescence in the intense electric field in the liquid inside the holes; successive amplification of the resulting UV photons in the cascaded structure could then result in detectable photon (and possibly charge) signals [1].

Preliminary experiments [7] showed large light yields emitted from of a THGEM electrode immersed in liquid xenon (LXe). Further systematic studies [8], however, provided strong – but *indirect* – evidence that the mechanism was in fact electroluminescence in xenon bubbles trapped below the electrode.

Here, we report on the direct visual observation of a xenon gas layer formed in LXe under a THGEM electrode and on the correlation of its appearance with radiation-induced electroluminescence signals. The process was shown to be stable over several days, providing an energy resolution of $\sigma/E \approx 7.5\%$ for ionization electrons extracted from 5.4 MeV alpha-particle tracks (~6,000 deposited ionization electrons per event).

## 2 Experimental setup and procedures

*Cryogenics and gas handling system*

The experiments were conducted in a new cryogenic setup (MiniX) - a small-volume system that permits fast (two-day) turnaround from opening to data taking. The system (Figure 1) comprises a Ø4" stainless steel vacuum-insulated chamber of ~0.5 liter LXe capacity. Xenon is liquefied with liquid nitrogen ($LN_2$), via a copper ring soldered to the chamber circumference; the ring is coupled to a 500 mm-long Ø1.5" copper rod immersed in the $LN_2$. Two 50 Ω heater connected to a cryogenic temperature controller (Cryo-con Model 24C) are used to stabilize the ring's temperature within ~0.1K. Xenon liquefaction occurs at the coldest part of the stainless steel chamber, which is at the copper ring location (100 mm above the chamber bottom and 200 mm below the chamber top flange). A CF-flanged upper turret houses all service ports and feedthroughs. A PTFE cylinder equipped with a Pt100 temperature sensor is placed at the bottom of the chamber within the LXe in order to reduce the amount of unused Lxe; the PTFE structure also assures regular flow of the liquid condensed at the copper ring to the chamber's bottom. A photomultiplier tube (PMT, Hamamatsu R8520) is placed at the bottom in another PTFE block, facing upwards. A slightly tilted Suprasil quartz window, placed above the PMT, deflects residual uncontrolled bubbles to the side so that they do not reach the THGEM region. The inner PTFE block is also equipped with two white LEDs to illuminate the inner-chamber volume.

Xenon purification is done by pumping the liquid out of the cryostat, through a homemade coaxial countercurrent heat exchanger. LXe evaporates in the heat exchanger after which the gas is compressed with a double-diaphragm recirculation pump (ADI R016-FN-CB2-D), whose output is regulated by a mass flow controller (Aalborg GFC17S-VCL2-AO, typical flow of 2 slpm). The gas then flows through a purifier (hot getter, SAES PS3-MT3-R-2); from whence it returns to the chamber through the heat exchanger. All gas system components (except the recirculation pump) are UHV-rated. A 500 liter aluminum tank is connected to the system through a pressure relief valve, for Xe recuperation in case of an accidental pressure rise.

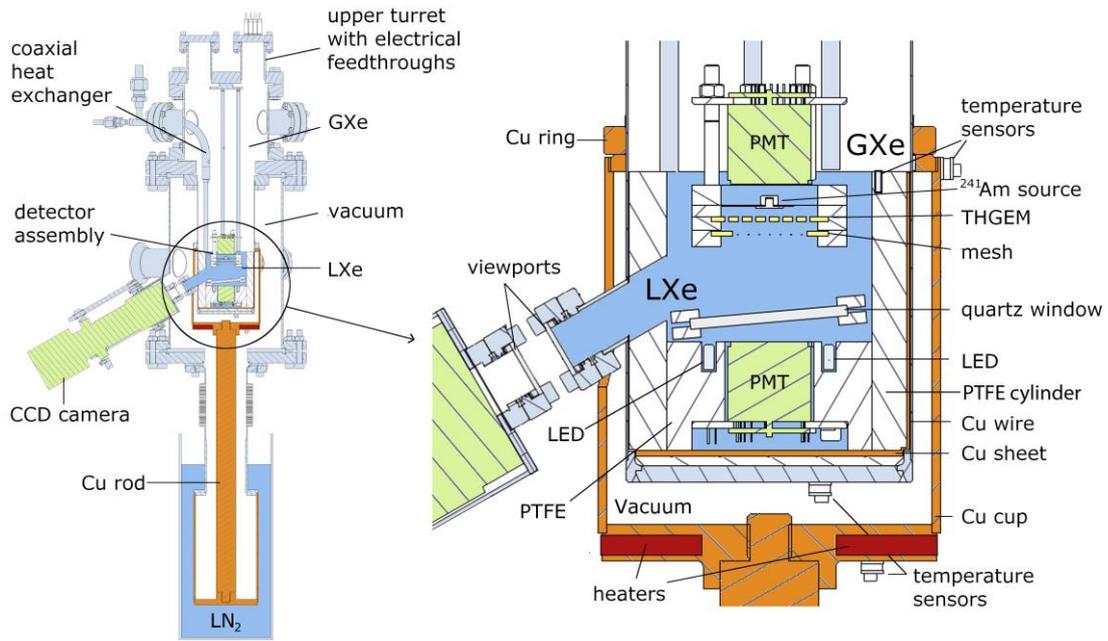

Figure 1 Left: Cross section view (to scale) of the MiniX cryostat. Right: magnification of the liquid xenon chamber, showing the cooling mechanism, the location of the temperature sensors and heaters, the internal PTFE block and cylinder and the detector assembly.

*Detector Prototype*

The detector prototype is installed in a PTFE housing, suspended by four stainless-steel rods from the upper DN160 CF8" top flange (Figure 1). The flange has all the necessary feedthroughs for the detector operation, allowing for its rapid assembly and installation in the cryostat.

The detector assembly is modular; it was designed for studying the operation and properties of different electrode configurations, for the direct observation and control of bubble formation and for the detection of electroluminescence signals. It is comprised of a stack of PTFE electrode holders, a second PMT (Hamamatsu R8520) at the top facing down, an $^{241}$Am alpha source (activity of ~180 Bq) deposited on an electrically-biased stainless steel holder and a THGEM electrode (0.4 mm-thick FR4 THGEM Cu-clad and Au-plated on both faces, with a triangular matrix of Ø0.3 mm holes, 1 mm-pitch and 0.1 mm-wide etched hole rims). For the current study, an electrode made of a grating of Ø55 μm NiFe resistive heating wires spaced 2.4 mm apart (resistance ~1.80 Ω/cm), was located 4.5 mm under the THGEM electrode. The wire electrode served to define the electric field below the THGEM. Only the central wire was used for heating; the application of a current through this wire generated bubbles under the THGEM (see details in Figure 2).

Alpha particles emitted into the liquid produce both scintillation photons, detected mainly by the top PMT via reflections from the PTFE-holder ("S1" signal) and ionization electrons; the latter drift under the application of an electric field towards the THGEM and eventually induce electroluminescence ("S2" signal). S2 signals (and a small fraction of the S1 photons) are recorded from the bottom PMT. The PTFE holders were designed such that a bubble created under the THGEM would remain in place (Figure 2).

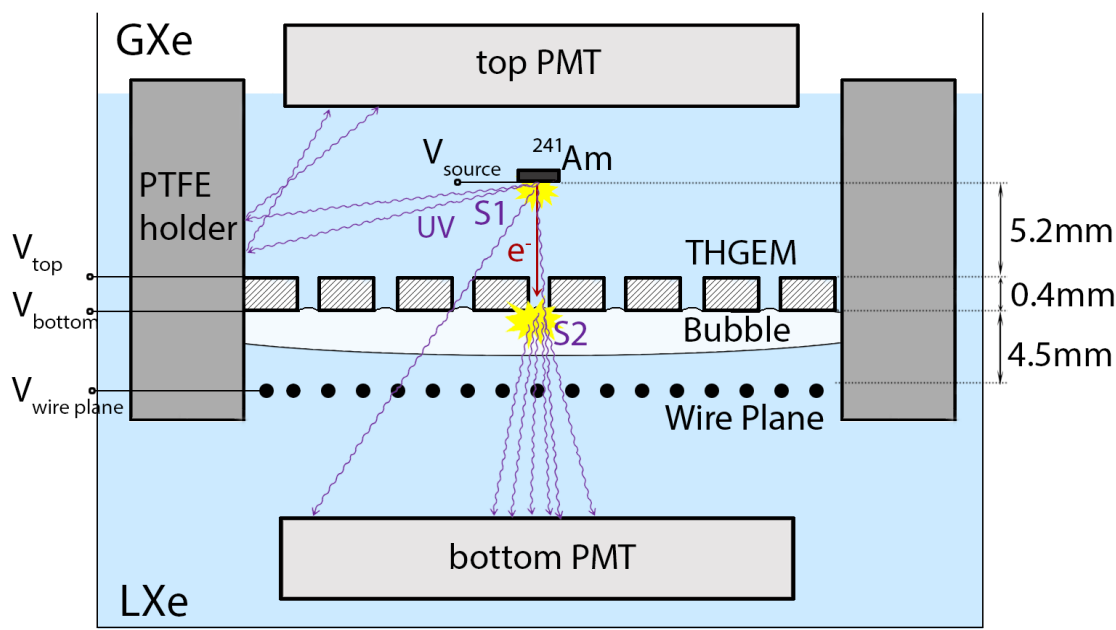

Figure 2. Schematic view of the detector assembly, immersed in LXe. A PTFE structure is holding the top PMT, an $^{241}$Am alpha source, a THGEM electrode and a wire plane electrode; in the latter the middle resistive wire can be heated to generate bubbles. A large bubble is formed, supported by the bottom part of the THGEM electrode; it is laterally contained within the PTFE walls. Alpha particles emitted into the liquid induce both S1 photons, recorded mainly by the top PMT due to reflections, and ionization electrons; the latter drift towards the THGEM electrodes' holes where they are extracted into the bubble and generate the S2 electroluminescence signal - recorded mainly by the bottom PMT.

*Viewing the detector chamber*

The cryostat system has two windows into the xenon volume. A single DN40 CF2.75" viewport is located on the top flange, providing a view from above. Two coaxial DN16 CF1.33" glass viewports (between the LXe and the insulation vacuum and between vacuum and air), located on an axis 60° with respect to the chamber's axis (see Figure 1), permit viewing the detector from below. Images are recorded with a CCD camera (CALTEX VIP-50-HD60) with optics allowing magnification of up to ×50 at a relatively large working distance (~200 mm). Illumination is provided by the two white LEDs mentioned above.

*Cooling down and liquefaction*

Prior to liquefaction, the detector chamber was evacuated with a dry turbo-molecular pump, to vacuum levels of $10^{-5}$ mbar. This was followed by Xe filling through the SAES getter and gas recirculation for at least 2 hours at room temperature, before liquefaction. Liquefaction of ~0.25 liters occurs within ~3 hours after the beginning of the cool down. Experiments were carried out at typical LXe temperatures of 165-175 K and pressures in the 1-2 bar range. S2 signals, when observed, were recorded after LXe recirculation for at least 12 hours at 2 slpm. Experiments under controlled bubble formation were carried out once the system reached steady state (at least 12 hours after liquefaction), both in liquid recirculation mode and without circulation. Vacuum insulation is kept below $5\times10^{-4}$ mbar using a turbo-molecular pump. As long as the cold rod is immersed in LN$_2$, vacuum levels below $5\times10^{-4}$ mbar can be maintained without continuous pumping.

*Electroluminescence signal recording*

In the present setup, bubble photography could not be done simultaneously with S2 signal recording from the bottom PMT, because of the light from the LEDs. Therefore, correlated bubble pictures and S2 events were recorded at time lapses of a few tens of seconds. This is acceptable because of the long-term stability of the bubbles formed under the THGEM electrode (from hours to days).

# 3    Results

## 3.1    Spontaneous bubbles formation

The first Xe liquefaction cycles in MiniX showed spontaneous generation of bubbles in the liquid. The bubbles appeared without applying current to the heating wire or potentials to the various inner elements. Thus, before proceeding with controlled experiments, efforts were made to understand and minimize the sources for these bubbles.

A major source of boiling was found to be heat conduction down the wires used to bias the electrodes and supplying power to the PMTs and LEDs. The wires were mostly Accuglass 100670 Ag-plated Cu with polyimide insulation (conductor Ø 250 μm; overall Ø 500 μm). In a first experiment, a single wire of the same type was coupled to one of the upper-flange connectors (kept at room temperature); half of it traversed the Xe gas (whose temperature varied gradually from 293K on the top down to 170K at the liquid surface) and half was immersed in LXe, with ~2mm stripped from its insulation at the end. Figure 3 and Video 1 show constant bubble formation emanating from the end of the polyimide. In another experiment, an additional ~10mm of the insulation was removed just below the surface of the liquid (midway along the wire's length). This treatment completely eliminated boiling from the end of the wire, which enabled us to continue to use this wire where the high voltage rating of thick polyimide was essential. The boiling that occurred just below the liquid surface (but not visible to the camera), being far away from the detector's sensitive volume, no longer caused a problem. The rest of the wires were replaced by low thermal conductivity 36 AWG (Ø127μm), Formvar®-insulated phosphor-bronze wires (CryoCon PW4-36-100). Furthermore, a copper sheet was placed at the bottom of the chamber to which three vertical ~1.3 mm thick copper wires, ending under the liquid surface near the cooling ring, were soldered to improve the liquid temperature uniformity (see Figure 1). In addition, the detector assembly was shielded from the residual parasitic boiling (e.g. boiling caused by dissipation in the bottom-PMT base and/or by the LEDs), with a slightly tilted quartz window placed above the bottom PTFE block (as mentioned above, seen in Figure 1). The tilt prevented bubble accumulation above the bottom PMT, which could lead to uncontrolled photon loss and undesirable fluctuations due to reflections and refraction at the liquid-gas interface.

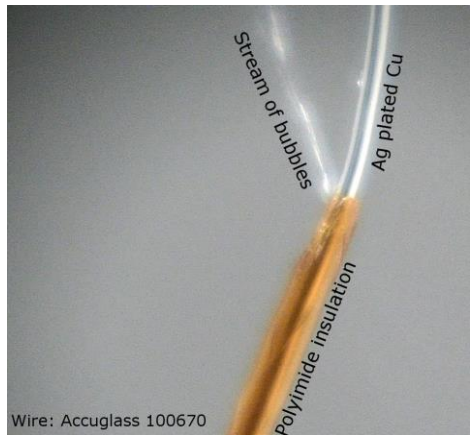

Figure 3  Boiling at the stripped end of a polyimide-insulated wire immersed in LXe as seen by the camera. The wire (of total length ~25cm) is coupled to an external connector, at room temperature. About half of the insulated wire's length traverses the Xe gas prior to its passage through LXe.

Since the inner parts of the chamber are warmer than the LXe at the beginning of every run, bubbles are created at these parts until they cool down. Each cycle of experiments began with these spontaneously-created bubbles trapped under the THGEM electrode. Similarly to [8], the bubbles could be eliminated by raising the pressure in the detector vessel; this was done either by adding Xe to the chamber or by starting (or increasing) the recirculation flow. The pressure increase would momentarily drive the system out of its steady state, causing bubbles to collapse. With the precautions taken above, occasional residual bubbles appeared mostly outside of the sensitive volume. In most runs, after having taken all the above steps and after letting the chamber's temperature stabilize for more than 12 hours, there were no bubbles left visible in the detector's volume.

### 3.2  Controlled bubble generation

Figure 4 presents photos of four typical conditions observed below the THGEM. The upper white part in each image is the PTFE detector holder. To the right of each photo is a drawing depicting the same conditions: A) THGEM electrode immersed in the liquid phase; B) liquid phase with small bubbles forming on the heating wire (under applied current); C) liquid phase with a fraction of a large bubble (heating wire current turned off); D) liquid phase with a large bubble covering the whole surface (heating wire current turned off). Note that since the camera views the THGEM at an angle larger than that of total internal reflection from the liquid-gas interface (45°-46° at visible wavelengths [9]), one can observe in C) and manifestly in D) both the wires and their reflected image. The entire bubble formation process is shown in Video 2.

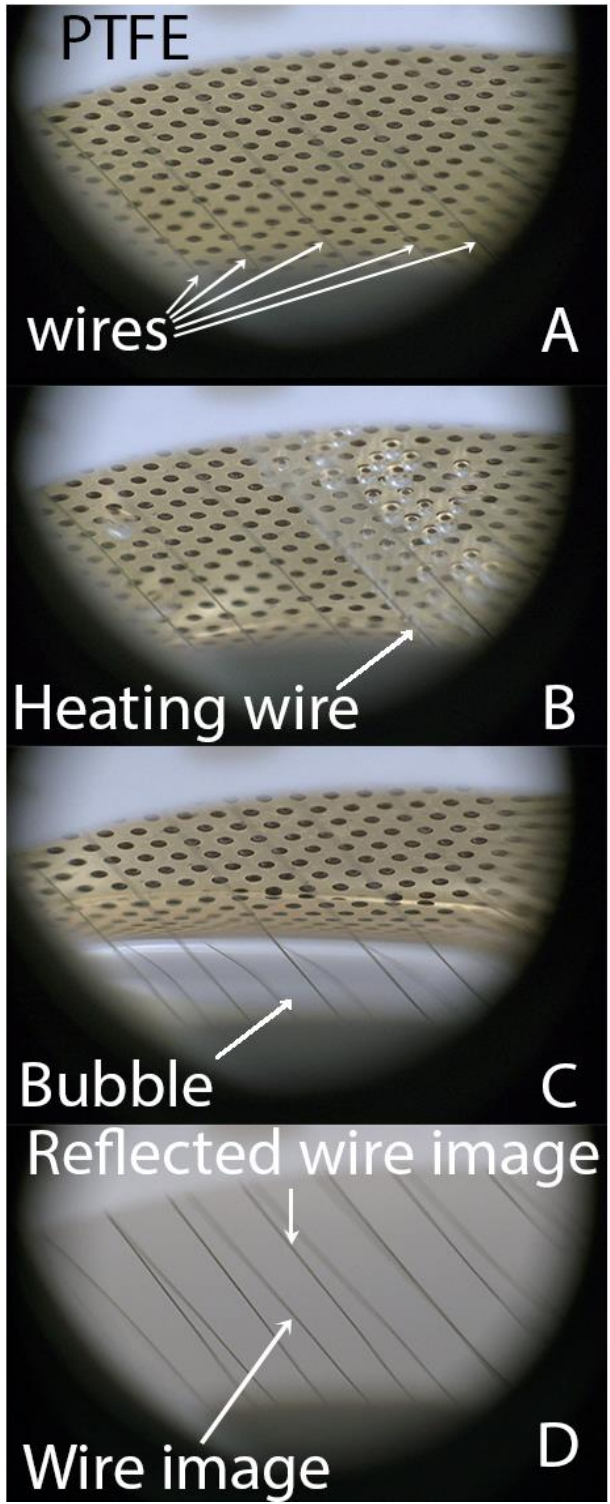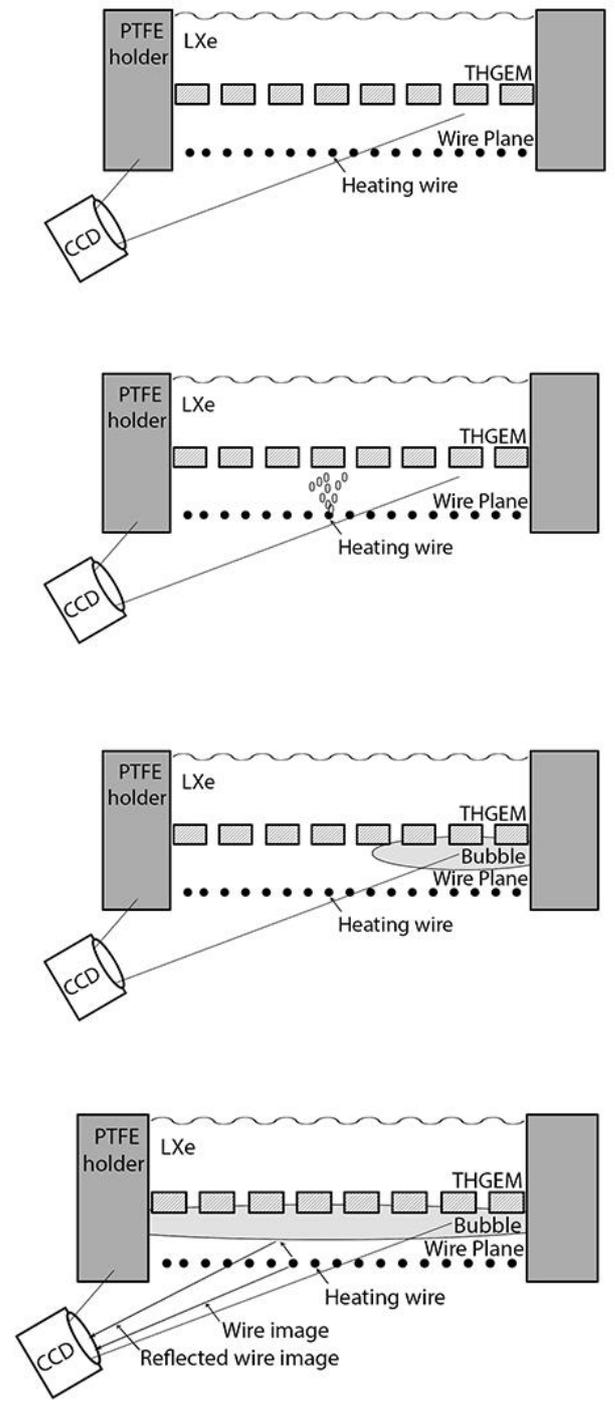

Figure 4 Views into the xenon volume as seen with the camera: A) the THGEM immersed in liquid, prior to bubble formation; visible are the THGEM bottom, the wires and a part of the PTFE holder (white). B) bubbles generated in the LXe by the heated wire trapped under the THGEM. C) A large-bubble covering a significant fraction of the THGEM underside, formed by the coalescence of numerous small bubbles. D) a bubble covering the entire area below the THGEM; both the wires and their reflections from the liquid-gas interface are seen.

### 3.3 Bubble formation and its dynamics

The intentional formation of a bubble under the THGEM electrode required the application of currents >300 mA through the heating wire. At lower currents, the heat was carried away by convection in the liquid surrounding the wire, and was clearly visible. Bubbles generated around the wire were typically sub-mm in size; they detached from the wire, reaching the bottom surface of the THGEM, where they coalesced within a few minutes into a single large bubble, an example of which is shown in Figure 4D. Once the bubble reached the dimensions of the surrounding PTFE holder, further heating of the liquid resulted in an increase of its thickness.

Upon the formation of a Xe bubble large enough to cover the entire THGEM area (~4 seconds at ~355 mA), the heater was turned off and the dynamics of the bubble were observed for some tens of minutes. It was observed that the bubble was slightly 'breathing' namely, that its thickness varied with a period of some tens of seconds (see Video 3). Occasionally, the bubble would shrink to a point where it no longer covered the entire surface of the THGEM; sometimes it extended beyond the wire plane. Since this was observed after the system reached a state with no spontaneous bubbling for several hours, the 'breathing' is probably not due to residual boiling. We further noted that heating the liquid without additional bubbling (at ~250 mA) seemed to improve the stability of the existing bubble by extending its breathing period and reducing the variations in its thickness. This phenomenon is under investigation. The bubble remained in place (without additional heating) for a period of a few days until the experiment was stopped.

### 3.4 Correlation of S2 signals with bubbles

As mentioned above, the recording of electroluminescence signals with the bottom PMT and bubble photography were performed alternately (at intervals of a few tens of seconds). The electrodes (THGEM, source and wire plane) were biased to provide electroluminescence yields similar to that observed in [8] and [7]. The values of the different potentials applied to the electrodes in the setup shown in Figure 2 and for which the results are presented here were the following: $V_{source}$ =-3100 V, $V_{top}$=-2700 V, $V_{bottom}$ =-300 V, $V_{wire\ plane}$=0 V (corresponding to the field at center of the THGEM holes of ~43 kV/cm, to a drift field above the THGEM $E_{drift} \approx 0.8$ kV/cm, and to a transfer field below the THGEM $E_{transfer} \approx 0.7$ kV/cm). These values provide complete collection of electrons into the THGEM holes (see Figure 10 in [8]). COMSOL simulation of the electric field inside the holes showed that at the liquid-gas interface, the magnitude of the electric field is >10 kV/cm; it therefore provides full extraction of the electrons from the liquid to the gas phase [2]. The top PMT, which provides the trigger on S1 signals, was operated at -700 V; the bottom PMT was operated at -600V to avoid pulse saturation.

Initially, prior to bubble formation, no electroluminescence signals were observed. With sufficient current flowing in the heating wire, as described above, a large bubble formed under the whole surface of the THGEM electrode, as described above (Figure 4D). Once stable for ~30 minutes, the illumination was turned off and the PMT was turned on – recording the S2 signals. Repeating the process of creation and annihilation of the bubble, while alternately observing S2 signals and the camera picture, proved that whenever S2 signals were visible, there was a bubble under the THGEM. Inversely, no S2 pulses were detected in the absence of the bubble.

### 3.5   S2 resolution

Typical S1 and S2 pulses recorded from the bottom PMT are shown in Figure 5. A spectrum of the alpha-particle-induced S1 scintillation signal recorded from the top PMT is given in Figure 6A; the main peak corresponds to 5.4 MeV alpha particles, the tail to the left corresponds to particles emitted from the source at an angle leaving a fraction of their energy in the sources matrix. The S2 spectrum shown in Figure 6B was recorded from the bottom PMT - triggered by S1 from the top PMT. The S1 trigger resulted in the recording of only alpha-particle-induced events (including coincident ones of alpha-particle and 59.5keV gamma-rays also emitted by the $^{241}$Am source - as discussed in [8]). The left tail corresponds to the lower-energy alpha particles as described for the S1. The excess of events to the right of the peak corresponds to a fraction of the alpha-gamma coincidence events not removed by the analysis (for further details see [8]). A series of ~14,000 waveforms were recorded per experiment on an oscilloscope (Tektronix TDS5054B) during ~10 min and were post-processed by dedicated Matlab scripts. Waveform analysis excluded most alpha- + gamma-induced events (for details see [8]). A Gaussian fit was performed on the main peak from which the mean S2 value and the resolution were extracted. The S2 resolution, in the present controlled bubble formation regime, reached values as high as $\sigma/E \approx 7.5\%$; it is superior to that reached previously in the 'super stable S2' regime in [8] (11.5%), under spontaneous bubble generation.

The number of electrons per alpha-particle event, reaching the THGEM and inducing the S2 electroluminescence signal in the bubble underneath (assuming their full collection into the holes [8]) was measured using a low-noise charge sensitive preamplifier (Canberra 2006). The charge signals were digitized and averaged over 1,000 waveforms. For drift field values of 0.9kV/cm, the charge collected onto the THGEM electrode was ~1 fC, i.e., equivalent to ~6,000 electrons.

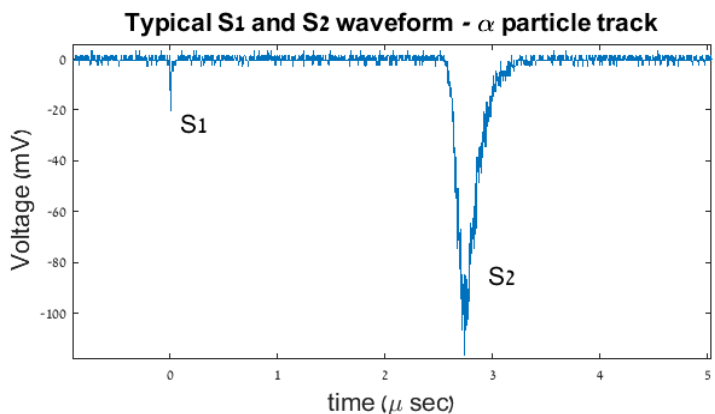

Figure 5 Typical S1-S2 pair as seen on the oscilloscope directly from the bottom PMT.

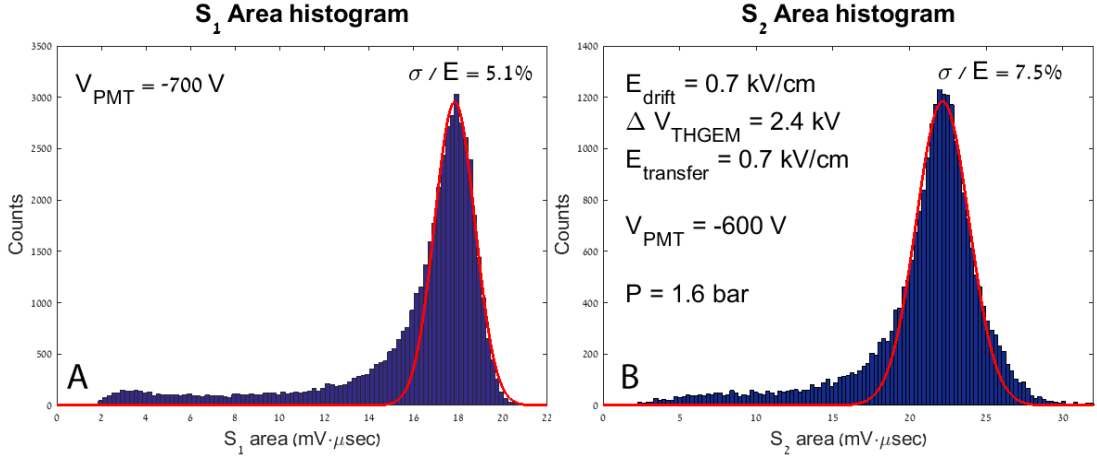

Figure 6 A) Pulse-area histograms of the S1 scintillation signal as recorded with the top PMT. B) the electroluminescence signal S2 recorded with the bottom PMT (setup of Figure 2). The red lines indicate a Gaussian fit from which the mean value and the resolution were calculated.

## 4   Summary and discussion

The primary result of the present work is the visual confirmation of the indirect evidence given in [8], that the electroluminescence signal observed from THGEM holes immersed in LXe occurs in a gas layer trapped under the electrode. This gas layer, created either spontaneously (e.g. through heat conduction of system components) or in a controlled way by locally heating the liquid, can by sustained for long time periods. This operation mode provided energy resolution as high as $\sigma/E \approx 7.5\%$ for alpha-particle-induced charge of ~ 6,000 electrons.

In order to compare our result to reported S2 resolution in other experiments using dual-phase Xe TPCs, one needs to refer to events having similar deposited charge. For example, a signal of 6,000 electrons measured in our experiment would correspond to that of a gamma-photon with energy of ~120 keV interacting in the XENON100 detector [4]: for a drift field of 0.53 kV/cm (as in XENON100), a charge yield of ~50 electrons/keV is expected [2]. The σ/E values for S2 reported in [4] are 12±1% for 164 keV and 14±1% for 80 keV gamma-rays.

It is worth mentioning that this work points out the problem of spontaneous bubble formation of LXe; it could be a potential source of operational problems in large-volume noble-liquid detectors. In addition to a possible cause for discharges, gas bubbles could affect the stability of the liquid-vapor interface in dual phase detectors – and thus their resulting energy resolution.

While this work confirms the possibility of controlling reproducible bubble formation and stability over relatively long periods, further studies are ongoing to evaluate the cascaded-LHM concept [1]. Other studies are planned to evaluate the idea of recording the electroluminescence photons by Gaseous Photomultipliers (GPM [10]). The potential implication of the LHM concept on the design of single-phase TPCs for future large-scale experiments (dark matter, neutrino physics) is yet unknown and requires further intensive study.

**Acknowledgements**

This work was partly supported by the Minerva Foundation with funding from the German Ministry for Education and Research (Grant No. 710827) and the Israel Science Foundation (Grant No. 477/10). We thank B. Pasmantirer and O. Diner from the Design Office and S. Assayag and members of his Mechanical Workshop at the Weizmann Institute, for their invaluable assistance in the design and manufacture of the experimental MiniX setup; we also thank A. Kisch of Zurich University for the manufacture of the PMT bases and O. Aviv, L. Broshi, Z. Yugrais and T. Reimer from Soreq NRC for the preparation of the $^{241}$Am source. The research has been carried out within the DARWIN Consortium for future dark matter experiments. A. Breskin is the W.P. Reuther Professor of Research in The Peaceful Use of Atomic Energy.